\documentclass[twocolumn,english,final,showpacs,superscriptaddress,lengthcheck,prl]{revtex4}
\usepackage[T1]{fontenc}
\usepackage[latin9]{inputenc}
\usepackage{textcomp}
\usepackage{amsmath}
\usepackage{graphicx}
\usepackage{amssymb}

\makeatletter
\@ifundefined{textcolor}{}
{%
 \definecolor{BLACK}{gray}{0}
 \definecolor{WHITE}{gray}{1}
 \definecolor{RED}{rgb}{1,0,0}
 \definecolor{GREEN}{rgb}{0,1,0}
 \definecolor{BLUE}{rgb}{0,0,1}
 \definecolor{CYAN}{cmyk}{1,0,0,0}
 \definecolor{MAGENTA}{cmyk}{0,1,0,0}
 \definecolor{YELLOW}{cmyk}{0,0,1,0}
 }

\@ifundefined{definecolor}
 {\@ifundefined{definecolor}
 {\@ifundefined{definecolor}
 {\usepackage{color}}{}
}{}
}{}
\usepackage{subfigure}  
\usepackage{hyperref}   

\makeatother

\usepackage{babel}

\makeatother

\usepackage{babel}

\makeatother

\usepackage{babel}

\begin{document}

\title{Zeno and anti-Zeno polarization control of spin-ensembles by induced
dephasing}

\author{Gonzalo A. \'{A}lvarez}

\affiliation{Facultad de Matem\'{a}tica, Astronom\'{\i}a y F\'{\i}sica, Universidad
Nacional de C\'{o}rdoba, 5000 C\'{o}rdoba, Argentina.}

\affiliation{Fakult\"{a}t Physik, Universit\"{a}t Dortmund, Otto-Hahn-Strasse
4, D-44221 Dortmund, Germany.}

\author{D. D. Bhaktavatsala Rao}

\affiliation{Department of Chemical Physics, Weizmann Institute of Science, Rehovot,
76100, Israel.}

\author{Lucio Frydman}

\email{Lucio.Frydman@weizmann.ac.il}

\affiliation{Department of Chemical Physics, Weizmann Institute of Science, Rehovot,
76100, Israel.}

\author{Gershon Kurizki}

\email{Gershon.Kurizki@weizmann.ac.il}

\affiliation{Department of Chemical Physics, Weizmann Institute of Science, Rehovot,
76100, Israel.}


\begin{abstract}
We experimentally and theoretically demonstrate the purity (polarization)
control of qubits entangled with multiple spins, using induced dephasing
in nuclear magnetic resonance (NMR) setups to simulate repeated quantum
measurements. We show that one may steer the qubit ensemble towards
a quasi-equilibrium state of certain purity, by choosing suitable
time intervals between dephasing operations. These results demonstrate
that repeated dephasing at intervals associated with the anti-Zeno
regime lead to ensemble purification, whereas those associated with
the Zeno regime lead to ensemble mixing. 
\end{abstract}
\maketitle
\emph{Introduction.---} The ability to understand and manipulate the
dynamics of {}``open'' quantum systems, \textit{i.e.} systems that
interact with their environment ({}``bath''), is a major challenge
for fundamental quantum physics and a prerequisite for novel applications
such as quantum heat engines \cite{scu}, quantum information storage
and retrieval \cite{molmer} and precision measurements \cite{prec}.
This is particularly true of information-carrying spin-$1/2$ particles,
known as quantum bits (qubits), coupled to spin-$1/2$ particles of
other species that constitute the bath. Such systems are usually controllable
by coherent fields \cite{viola,goren,natexp2,Ernst}. Here we address
their manipulations by incoherent, random fields that dephase the
system and thereby mimic quantum non-demolition (QND) measurements
\cite{qnd,prl87}. Such manipulation is conceptually intriguing: whereas
QND measurements leave a closed system intact, they can affect an
open system, by destroying its correlations (coherences) with the
bath. As recently predicted for qubits coupled to thermal oscillator
baths, such measurements can steer the qubit ensemble, towards either
higher or lower purity ({}``cooling'' or {}``heating'') \cite{Noam08}:
Namely, the qubit does not retain its state as the measurements accumulate,
but rather converges to an asymptotic steady state. In the Quantum
Zeno (QZE) regime, highly frequent measurements raise the asymptotic
excitation and entropy (mixing) of the qubit. This reflects the hitherto
unnoticed fact that QZE dynamics equalize the bath-induced upward
and downward transition rates, in the qubit. By contrast, less frequent
measurements conforming to the anti-Zeno (AZE) regime \cite{Noam08},
predominantly enhance downward transitions (relaxation to the ground
state) and thus purify ({}``cool'') the qubit. These predicted measurement-induced
changes of the equilibrium go beyond previous studies that focused
on transition-rate (relaxation) slowdown in the QZE regime and its
speedup in the AZE regime \cite{prl87} and have been experimentally
verified \cite{exps}. %

The present study considers a scenario different from that of Ref.
\cite{Noam08}: the interaction of a spin-$1/2$ or qubit system $S$
with $N$ identical spin-$1/2$ particles $I$ that constitute its
{}``bath''. Such a situation is encountered in NMR setups \cite{Ernst},
and field-driven quantum dots \cite{qdot1}. Since all the $I$ spins
have the same energy levels, such spin-baths are spectrally degenerate,
as oposed to the broad spectrum of oscillator baths. The resulting
qubit-bath dynamics is therefore different for the two scenarios and
hence we ask: do the equilibrium changes predicted in Ref. \cite{Noam08}
hold for both scenarios?

In this work we demonstrate that they do, despite their differences:
we experimentally lower or raise the purity of the system and bath
spins via frequent induced dephasings that simulate QND energy measurements
\cite{prl87} respectively, by timing the dephasing intervals to be
in the QZE (evolution slowdown) or AZE (evolution speedup) regime
\cite{prl87} respectively. Remarkably, repeated dephasings at intervals
conforming to the AZE lead to highly-effective polarization exchange
that can overcome even large frequency detunings (off-resonant mismatch)
of the qubit and bath spins and induce polarization transfer that
is as large as if it were the Hartmann-Hahn resonant transfer \cite{HH62}.
This novel effect, demonstrated by our NMR polarization transfer experiments,
is termed here \textit{incoherent resonance}, as it stems from repeated
system-bath correlation erasure (dephasing).

\emph{Model and dynamical regimes.---} We assume that the qubit-bath
system is described by an effective Hamiltonian \begin{eqnarray}
H & = & H_{0}+H_{SI}+H_{M}(t).\label{eq1a}\end{eqnarray}
 having the following terms: (i) The $H_{0}$ Hamiltonian accounts
for the coherent evolution of the qubit and the bath; under a Zeeman-like
interaction with respective Larmor frequencies $\omega_{S}$ and $\omega_{I}$.
(ii) The $H_{SI}$ term describes the coupling between the $S$ and
$I$ spins, chosen for simplicity to be oriented perpendicular to
the Zeeman field, \begin{eqnarray}
H_{SI}=J\sum_{k}S^{x}I_{k}^{x},\label{eq1b}\end{eqnarray}
 $S^{x}$ and $I_{k}^{x}$ being the $\sigma_{x}$ Pauli operators
for the respective species. (iii) The time-dependent Hamiltonian $H_{M}(t)$
involves intermittently switched random fields that mimic repeated
QND measurements.

The incoherent $S$-$I$ cross-polarization transfer that we here
discuss is then determined by the interplay between {}``free'' evolution
and measurement effects, as follows:

\noindent \emph{(a) Free-evolution:} This will be governed by the
time-independent terms in Eq. (\ref{eq1a}). In an interaction picture,
\textit{i.e.,} in a {}``doubly'' rotating frame with frequencies
$\omega_{S}$ and $\omega_{I}$, $H_{SI}$ will have contributions
from both rotating-wave ($RW$ or flip-flop) terms $S^{+}I_{k}^{-}$,
oscillating as ${\rm e}^{it(\omega_{S}-\omega_{I})}$, and co-rotating
($CR$ or flip-flip) terms $S^{+}I_{k}^{+}$ oscillating as ${\rm e}^{it(\omega_{S}+\omega_{I})}$,
and by their respective Hermitian conjugate terms. It is clear that
the short-time initial evolution will be dominated by the rapidly
oscillating $CR$ terms, and the long-time evolution by their $RW$
counterparts. The energy transfer from $I$ to $S$ due to $CR$ and
$RW$ terms is then governed by the respective population transfer
coefficients \cite{EPAPS} $P_{CR}=\frac{\widetilde{J}^{2}}{\tilde{J}^{2}+(\omega_{S}+\omega_{I})^{2}}$,
$P_{RW}=\frac{\widetilde{J}^{2}}{\widetilde{J}^{2}+(\omega_{S}-\omega_{I})^{2}}$,
where $\tilde{J}$ is the effective $S-I$ interaction. At resonance
($\omega_{S}=\omega_{I}$), $P_{RW}=1$, causing a complete exchange
of polarization at $t\sim n\pi/J$ ($n=1,2,...$). In such situations,
and given that usually $\omega_{S},~\omega_{I}>J$, one can ignore
the fast-oscillating $CR$ terms and obtain the dynamics using the
$RW$ terms only \cite{HH62}. By contrast, under strongly mismatched
conditions, \textit{i.e.,} $\omega_{S}\gg\omega_{I}$ or vice-versa,
$P_{RW}\sim P_{CR}$, and the dynamics is equally dominated by the
$CR$ and $RW$ terms. For such large detuning, the $H_{SI}-$driven
transfer of polarization between the $S$ and $I$ spins is inhibited:
The polarizations of all spins are then locked at their initial values,
as $P_{CR}\sim J/|\omega_{S}+\omega_{I}|,P_{RW}\sim J/|\omega_{S}-\omega_{I}|\ll1$.
While the presented results are focused on the given $H_{SI}$, they
are general for Hamiltonians that contain $RW$ and $CR$ terms.

\noindent \emph{(b) Projective measurements:} These will be imparted
by brief interactions described by $H_{M}(t)$ \cite{EPAPS}. Each
such nonselective (unread) projective measurement \cite{Noam08} erases
the off-diagonal (correlations) terms that may have arisen in the
joint $S+I$ density matrix. This is equivalent to subjecting the
system to a brief strong dephasing. Although the respective eigenstates
of the system and the bath remain unchanged during these measurements,
their correlation energy $\langle H_{SI}\rangle$ changes drastically,
affecting subsequent evolution \cite{Noam08}. Here, we mimic such
projections onto the system's energy eigenbasis by an NMR {}``quantum
simulator'', \emph{i.e.}, spatially-random magnetic field-gradients
that changes over time (see below).

The polarization exchange between the $S$, $I$ spins is dramatically
altered in the presence of repeated projective measurements at times
$n\tau$, ($n=0,1,2,\cdots$). To appreciate this we consider initially
uncorrelated equilibrium states $\rho_{S}\otimes\rho_{I1}\otimes...\rho_{IN}$,
\emph{i.e.} products of the $2\times2$ density matrices of the $S-$system
and each of the $N$ $I-$bath spins that are diagonal in the energy
eigenbasis, with populations of the excited spin state being $0\le\epsilon_{S(I)}\le1/2$
(the corresponding polarizations $P_{S(I)}=1-2\epsilon_{S(I)}$).
We then find that $\epsilon_{S}(t)$ oscillates as the weighted sum
(over all possible $I$-spin quantum numbers) of $S-I$ oscillatory
exchange probabilities. This function depends on $N$, the bath size,
and the anisotropy of the spin ensemble \cite{EPAPS} but primarily
on the time between consecutive dephasings: (i) At very short times
$\omega_{S(I)}t\ll1$, the $S$ evolution is dominated by the fast-oscillating
$CR$ terms, so that the freely-evolving polarization of the $S$
spin is driven away from $\left(1-2\epsilon_{S}(0)\right)$, causing
depolarization of $S$ ({}``heating''): $\epsilon_{S}(t)<\epsilon_{S}(0)$.
This $CR$ {}``heating'' is amplified by the repeated QZE, since
$CR$ evolution dominates under the QZE condition $\left(\omega_{S}+\omega_{I}\right)\tau\ll1$
\cite{EPAPS}: {measurements or dephasing at intervals $\tau_{h}\le1/\sqrt{J^{2}+(\omega_{S}+\omega_{I})^{2}}$.
This condition means that highly frequent measurements broaden the
qubit levels to the extent that they become unresolved, equalizing
upward and downward transition rates regardless of temperature (Fig.
\ref{Fig1}a: red circles and Fig. \ref{Fig1}b: lower inset). (ii)
By contrast, at longer intervals, the $RW$ terms increase the polarization
(cause {}``cooling'') of $S$: $\epsilon_{S}(t)>\epsilon_{S}(0)$.
Such {}``cooling'', whose condition is $\left|\omega_{S}\pm\omega_{I}\right|\tau\gtrsim1$,
is amplified by the repeated AZE \cite{Noam08}: measurements or dephasing
at intervals $\tau_{c}\sim1/\sqrt{J^{2}+(\omega_{S}-\omega_{I})^{2}}$.
This comes about since at such intervals the qubit levels are resolved
and downward transitions (relaxation) dominate at finite temperature
(Fig. \ref{Fig1}a: blue upper triangles and Fig. \ref{Fig1}b: upper
inset). (iii) After a few measurements (see below) the polarization
transfer reaches close to the resonant maximum {[}$\epsilon_{I}(0)${]}
irrespective of the $S-I$ detuning. These time-scales determine a
resonant-like characteristic that can be exploited as shown in Fig.
\ref{Fig1}. The qubit polarization is then described, within the
$RW$ domain by Eq. (III.3) in EPAPS \cite{EPAPS}.

\noindent \emph{(c) Quasi-equilibrium:} After a few measurements at
suitable $\tau$'s, the polarization approaches its asymptotic value
and hence the system reaches a quasi equilibrium state (see below),
with polarization using RW terms only \begin{eqnarray}
\epsilon_{S}^{qe}=\epsilon_{S}(0)+\frac{\epsilon_{I}(0)-\epsilon_{S}(0)}{2(1-\epsilon_{I}(0))}.\label{eq5}\end{eqnarray}
 Depending on the sign of $\epsilon_{I}(0)-\epsilon_{S}(0)$, $\epsilon_{S}^{qe}$
can be either larger or smaller than $\epsilon_{S}(0)$, corresponding
to $S$-spin {}``cooling'' or {}``heating'', as compared to its
initial equilibrium value. The value of $(1-2\epsilon_{S}^{qe})$
is the largest obtainable polarization transfer from the $I$ spins
to the $S$ spin, for any size $N$ of the bath. 
The transfer achieved by the incoherent resonance is thus always greater
than $50\%$ of the coherent maximum, $\epsilon_{I}(0)/\epsilon_{S}(0)$,
and bound by the full coherent maximum obtainable under a resonant
transfer.

\noindent \emph{(d) Reheating:} Once $\epsilon_{S}^{qe}$ is reached,
the state of the total ($S+I$) system commutes with the interaction
Hamiltonian in the $RW$ approximation, $[\rho,H_{SI}]\approx0.$
This means that if no further measurements are performed, the evolution
of all the spins is almost frozen (Fig. \ref{Fig1}a: green lower
triangles). Yet, in a finite bath as measurements continue to be performed,
the deviations from Eq. (\ref{eq5}) due to the $CR$ terms, gradually
{}``re-heat'' (depolarize) both the $S$ and the $I$ spins (Fig.
1b: blue upper triangles). Hence, different desired quasi-equilibrium
values of the $S$-polarization can be obtained depending on $N$,
the bath size, and on the number of measurements performed beyond
the number needed to attain $\epsilon_{s}^{qe}$.

\begin{figure*}[ht]

{\includegraphics[trim=0 15 0 50,keepaspectratio,width=0.4\textwidth]{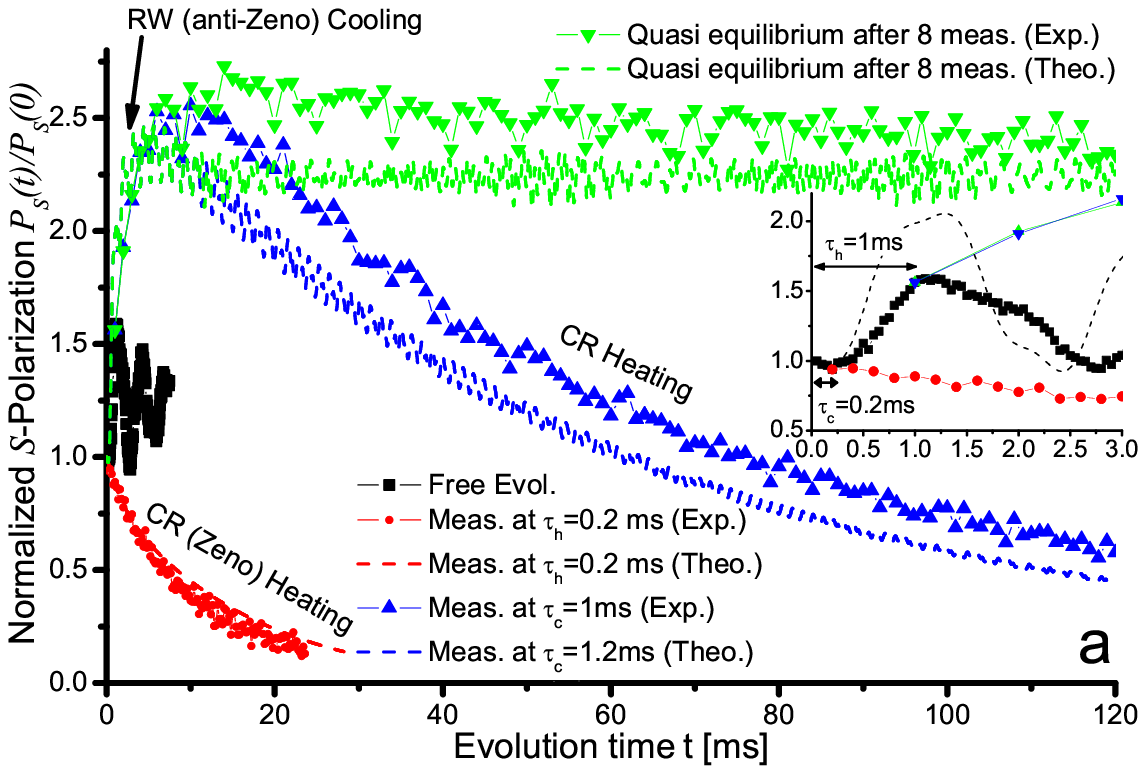}} {\includegraphics[keepaspectratio,width=0.35\textwidth]{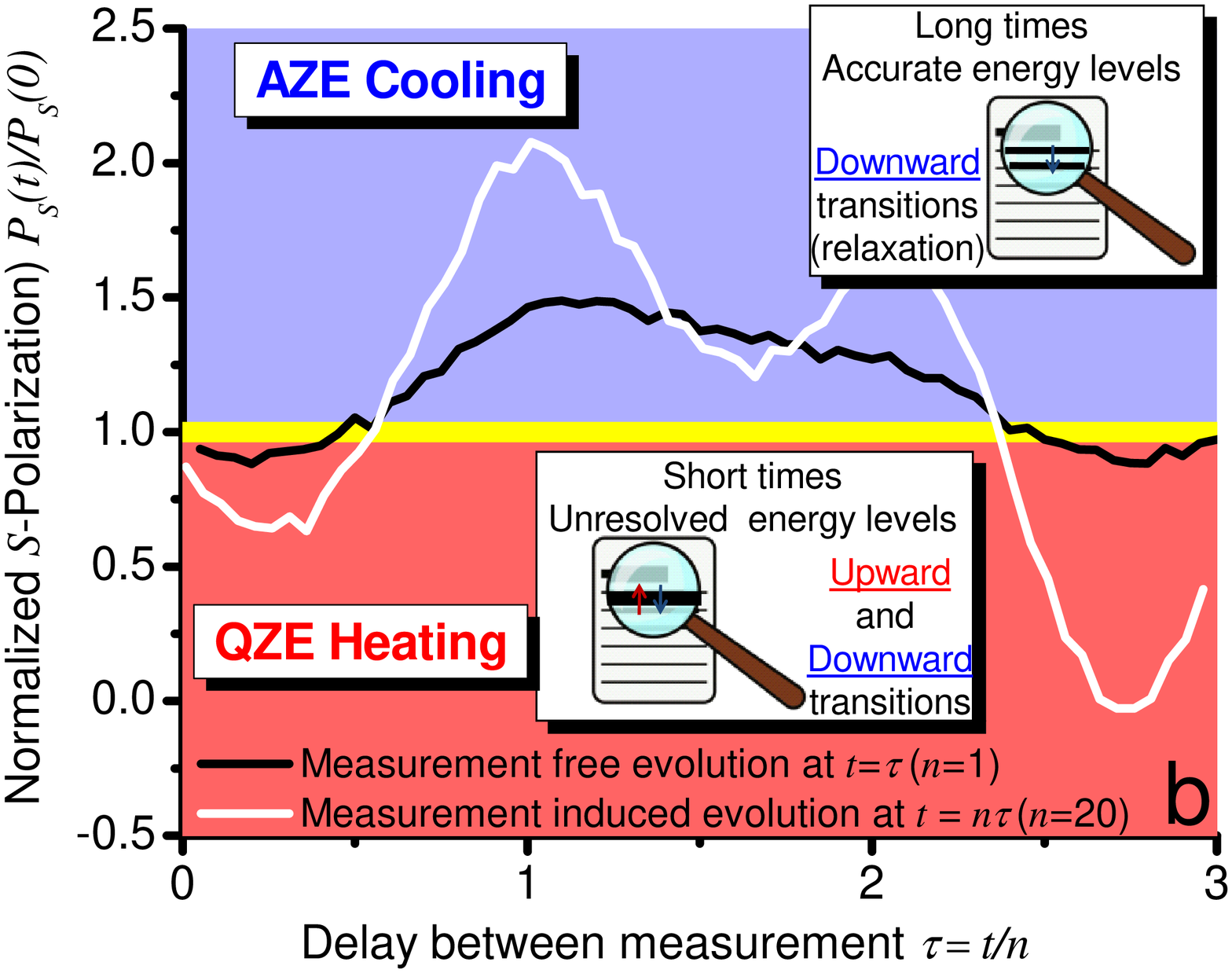}}

\caption{(Color online) Evolution of the $S$-spin polarization with time.
(a) The main panel compares the evolution of the $S$ spin polarization,
when interrupted by repeated measurements at intervals of $\tau_{c}^{exp}=1$
ms (blue upper triangles) and the evolution interrupted at $\tau_{h}=0.2$ms
(red circles), respectively. Also illustrated is the quasi-equilibrium
state achieved for the $\tau_{c}^{exp}$ measurements, stopped after
$8$ ms and followed by free evolution at later times (green lower
triangles). The free evolution of the $S$ spin is shown with black
squares. The inset zooms the dynamics for short times. These experimental
plots are compared with the theoretical curves (dashed lines) obtained
by exact diagonalization of the Hamiltonian in Eq. (\ref{eq1a}) for
the experimental parameters above. (b) Schematic representation of
the QZE and AZE in thermalized qubits. The white line is the quantum
dynamics of the $S$ spin steered by $n=20$ measurements for varying
the time interval $\tau$. It evidences the predicted amplifications
compared with the free evolution (black line). For short times (QZE
regime), the levels are unresolved so that their transition rates
are equal (lower inset), while for long times (AZE regime) they are
resolved so that the downward transitions dominate (upper inset). }

\label{Fig1} 
\end{figure*}

\emph{Results.---} The foregoing theoretical predictions which hold
for any size of the bath, were tested by liquid state NMR simulators
of QND measurement on $^{13}$C-methyliodide (CH$_{3}$I) dissolved
in CDCl${}_{3}$. In liquid-state NMR experiments on the methyl group
(CH$_{3}$) a $^{13}$C spin ($S$) is $J$-coupled to a finite {}``bath''
of $N=3$ equivalent $^{1}$H spins ($I$) which interact with the
$S$ spin but not with each other. The quasi-equilibrium value of
polarization obtained for $N=3$ is $\epsilon_{S}^{qe}=\epsilon_{S}(0)+\frac{1}{2}\left\lbrace (\epsilon_{I}(0)-\epsilon_{S}(0))[1+\epsilon_{I}(0)(1-\epsilon_{I}(0))]\right\rbrace $.
The Hamiltonian in Eq. (\ref{eq1a}) was reproduced by applying two
radio frequency (RF) fields perpendicular to the static field, on
resonance with the respective $I$ and $S$ spins. In the doubly rotating
frame of the RF fields, precessing with the respective Zeeman frequencies
of the spins we then obtain Hamiltonian (\ref{eq1a}) where the $z$
axis is given by the RF fields direction and the frequencies $\omega_{S}$
and $\omega_{I}$ determined by the strength of the respective RF
fields (see \cite{EPAPS} for details).

It is essential that both $T_{1}$ and $T_{2}$ relaxation times are
much longer than the time scales used for these quantum simulations.

To mimic the effects of projective measurements, we relied on the
use of pulsed magnetic field gradients. Field gradients effectively
increase the decoherence rate for correlations in a plane perpendicular
to the gradient's direction, which is along the main $B_{0}$ field
axis \cite{Jones06}. Yet, it can be shown that repeated application
of gradients gives rise to unwarranted temporal correlations in the
coherent evolution of the system \cite{EPAPS}. In order to realize
a sequence of dephasing operations, we resorted to an alternative
method that truly simulates projective measurements by applying at
regular intervals new, random, values of the field gradients \cite{EPAPS}.
In an ensemble average, the correlations are then effectively erased.

The initial conditions are $[1/2-\epsilon_{I}(0)]/[1/2-\epsilon_{S}(0)]\sim4$
with the excitations $[1/2-\epsilon_{I(S)}(0)]\ll1$. By choosing
the time intervals between measurements, to correspond to the QZE
or AZE regimes, depolarization ({}``heating'') or, respectively,
polarization ({}``cooling'') effects predicted by theory were indeed
observed under non-matching $\omega_{I}\sim250Hz$ and $\omega_{S}\sim420Hz$
(Fig. \ref{Fig1}a) where the choice $|\omega_{S}-\omega_{I}|$, $\omega_{S}$,
$\omega_{I}\sim J=150Hz$ enhances the predicted effects. Additional
measurements performed after attaining the maximum polarization caused
re-heating of the $S$ spin by $CR$ terms, as theoretically predicted
(blue upper triangles). Finally, by stopping the measurements after
maximizing the polarization transfer, we clearly observed the expected
quasi-equilibrium behavior of the $S$-spin ($^{13}$C) polarization
(green lower triangles). Its value agrees well with the theoretically
estimated $\left(1-2\epsilon_{S}^{qe}\right)\sim2.9\left[1-2\epsilon_{S}(0)\right]$
considering only the $RW$ term. Its slow decay is a consequence of
non-ideal pulses in the implementation of the projective measurements.
Excellent agreement is evident between experimental results and numerical
simulations without any fitting parameters. Fig. \ref{Fig1}b shows
the experimental $S$-polarization steered by $n=20$ measurements
as a function of their time interval $\tau$. Heating (purity loss)
or cooling (purity increase) are seen to depend on $\tau$ as predicted.
\begin{figure*}[ht]
 {\includegraphics[width=0.4\textwidth]{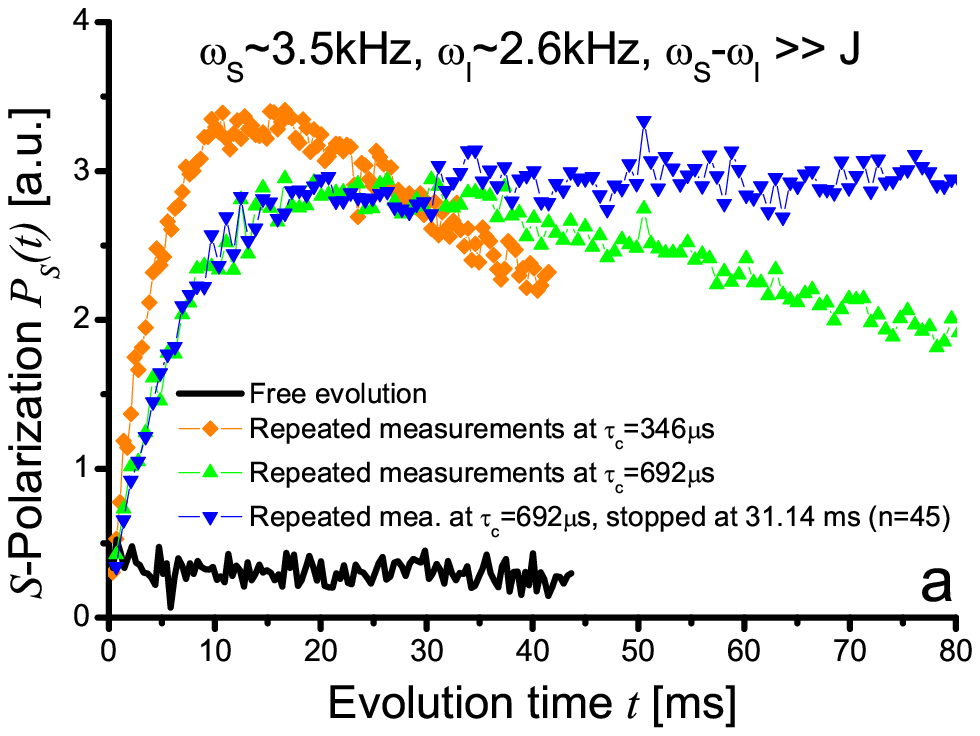}} {\includegraphics[width=0.4\textwidth]{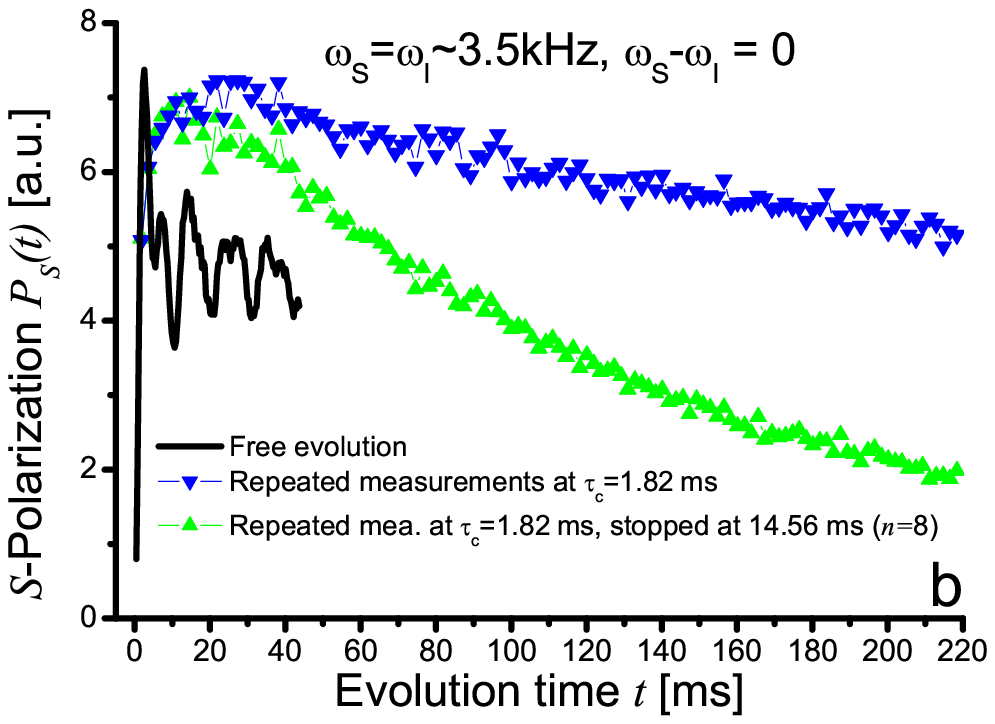}}

\caption{(Color online) Time evolution of the system polarization under matching
($\omega_{S}=\omega_{I}$) and off-matching ($\omega_{S}\ne\omega_{I}$)
conditions. (\textbf{a}). The free evolution (black solid line) of
the system ($S$) spin-polarization, and its evolution interrupted
by measurements at time intervals $\tau$ of $346\mu$s (orange circles)
and $692\mu$s (green upper-triangles) are shown for off-resonant
fields with high detuning. The given $\tau$ values were chossen to
optimize the transfer. The quasi-equilibrium state corresponding to
measurements (at time intervals of $692\mu$s) stopped after $31.14$
ms ($n=45$), followed by the free evolution for the later times is
marked by blue lower triangles. (\textbf{b}). For resonant RF fields,
we have plotted the free evolution (black line) of $S$ spin-polarization,
evolution interrupted by measurements at time intervals $1.82$ms
(green upper-triangle). The quasi-equilibrium state corresponding
to measurements (at time intervals of $1.82$ms) stopped after $14.56$
ms ($n=8$), followed by free evolution at later times, is marked
by blue lower triangles. The maximal polarization transfer attained
by resonant fields (black line) is almost the same as that achieved
and maintained (blue lower triangles) by measurements for all later
times. }

\label{Fig2} 
\end{figure*}

To further explore these incoherent polarization transfer effects,
larger detunings $|\omega_{S}-\omega_{I}|\gg J$ ($\omega_{S}\sim3.5{\rm kHz}$
and $\omega_{I}\sim2.6{\rm kHz}$) and fields $\omega_{S},\omega_{I}\gg J$
were probed and compared with the Hartmann-Hahn resonant \cite{HH62}
NMR transfer (Fig. \ref{Fig2}). In these experiments, only the $I$
spins were initially excited while the $S$ polarization was completely
erased {[}$\epsilon_{S}(0)=1/2${]} \cite{EPAPS}, so that the actual
polarization transferred from $I$ to $S$ spin could be determined.
Under these off-matching conditions, $P_{RW}$ and $P_{CR}$ were
much lower than their resonant values, and the transfer arising from
the {}``free'' (uninterrupted) dynamics was negligible (Fig. \ref{Fig2}a-solid
line). By contrast, the $I\rightarrow S$ polarization transfer achieved
by repeated projection measurements was unequivocally evidenced (Fig.
\ref{Fig2}(a)) to be close to the maximum achievable under resonant
conditions (Fig. \ref{Fig2}(b)).

\emph{Discussion.---} The theoretical and experimental data shown
in Figs. \ref{Fig1} and \ref{Fig2} clearly demonstrate for the first
time the connection between the fundamental Zeno and anti-Zeno effects
in frequently measured or dephased open systems and their purity loss
or increase, respectively. Measurement (dephasing)-induced transfer
of polarization has been shown to be almost as effective as coherent
on-resonance transfer, even if the measured system and bath are under
presumably unfavorable, off-resonant conditions. 
We have further demostrated the ability to steer the system into a
quasi-equilibrium, which is maintained when further measurements are
stopped.

In terms of their practical use, we envisage potential applications
of this non-unitary polarization transfer protocol for qubit purification,
required at the initialization stage of quantum information processing
\cite{divin}. The possibility of increasing the polarization transfer
from the pure $I$ spins to the impure $S$ spins even under unavoidable
off-resonant conditions could be useful for algorithmic cooling \cite{algo}.
Hence, the applied and fundamental aspects of the incoherent resonances
introduced here merit further exploration.

\begin{acknowledgments}
This work was supported by a Fundacion Antorchas-Weizmann Institute
Cooperation grant (\# 2530), and the generosity of the Perlman Family
Foundation. G.A.A. thanks CONICET and the Alexander von Humboldt Foundation
for financial support, and the hospitality of Fakult\"{a}t Physik,
TU Dortmund. Support by EC (MIDAS STREP), ISF, DIP and the Humbolt
Research Award is acknowledged by G.K. \\ 

\begin{thebibliography}{19}
\bibitem{scu} M. O. Scully, M.S.~Zubairy, G.S. Agarwal, and H. Walther,
Science \textbf{299}, 862 (2003).

\bibitem{molmer} J.H. Wesenberg, \emph{et. al.}, Phys. Rev. Lett.
\textbf{103}, 070502 (2009).

\bibitem{prec} V. Giovannetti, S.~Lloyd, and L.~Maccone, Science
\textbf{306}, 1330 (2004).

\bibitem{viola} L. Viola, E. Knill, and S. Lloyd, Phys. Rev. Lett.
\textbf{82}, 2417 (1999).

\bibitem{goren} G. Gordon, G. Kurizki, and D.A.~Lidar, Phys. Rev.
Lett. \textbf{101}, 010403 (2008).

\bibitem{natexp2} A.~Greilich, \emph{et. al.}, Nat. Phys. \textbf{5},
262 (2009).

\bibitem{Ernst} R.R.~Ernst, G.~Bodenhausen, and A.~Wokaun, {\em
Principles of Nuclear Magnetic Resonance in One and Two Dimensions},
(Clarendon, Oxford, 1987).

\bibitem{qnd} G. Gordon, \emph{et. al.}, J. Phys. B \textbf{40},
S61 (2007). V.B.~Braginsky and F.Ya.~Khalili, Rev. Mod. Phys. \textbf{68},
1 (1996).

\bibitem{prl87}A. G. Kofman and G. Kurizki, Phys. Rev. Lett. \textbf{87},
270405 (2001); Nature \textbf{405}, 546 (2000); Phys. Rev. Lett. \textbf{93},
130406 (2004).

\bibitem{Noam08} N.~Erez, \emph{et. al.}, Nature \textbf{452}, 724
(2008); G. Gordon, \emph{et. al.}, New J. Phys. 11, 12302 (2009);
G. Bensky, \emph{et. al.}, Physica E 42, 477 (2010).

\bibitem{exps} Q. Niu and M. G. Raizen, Phys. Rev. Lett. \textbf{80},
3491 (1998), F. Dreisow, \emph{et. al.}, Phys. Rev. Lett, \textbf{101},
143602 (2008).

\bibitem{qdot1} D. Loss and D. P. DiVincenzo, Phys. Rev. A \textbf{57},
120 (1998);P. Maletinsky, A. Badolato, and A. Imamoglu, Phys. Rev.
Lett. \textbf{99}, 056804 (2007).

\bibitem{HH62} S.R.~Hartmann and E.L.~Hahn, Phys. Rev. \textbf{128},
2042 (1962).

\bibitem{EPAPS}See EPAPS Document No. XXXXX for supplementary information.
For more information on EPAPS, see http://www.aip.org/pubservs/epaps.html.

\bibitem{Jones06} L.~Xiao and J.A.~Jones, Phys. Lett. A \textbf{359},
424 (2006).

\bibitem{divin} D.P.~DiVincenzo, Fortschr. Phys. \textbf{48}, 771
(2000).

\bibitem{algo} J.~Baugh, \emph{et. al.}, Nature \textbf{438}, 470
(2005). 

\end{thebibliography}

\begin{thebibliography}{19}
\bibitem{Ernst-1}C.P. Slichter, \emph{Principles of magnetic resonance},
2nd ed. (Springer-Verlag, 1992).

\bibitem{hbook}I.S. Gradshteyn and I.M. Ryzhik, \textit{Tables of
Integrals, Series, and Products} (Academic Press, USA, 1980). 

\end{thebibliography}
\end{acknowledgments}

\begin{widetext}

\newpage{}

\begin{center}
\setcounter{secnumdepth}{3} 
\setcounter{tocdepth}{10}
\def\thesection{\Roman{section}} 
\setcounter{MaxMatrixCols}{30}
\providecommand{\U}[1]{\protect\rule{.1in}{.1in}}
\numberwithin{equation}{section}
\def\be{\begin{equation}}
\def\ee{\end{equation}}
\def\e#1{\label{#1}\end{equation}}
\def\bea{\begin{eqnarray}}
\def\eea{\end{eqnarray}}
\def\nn{\nonumber}
\makeatother \textbf{\large Supplementary information: Zeno and anti-Zeno polarization
control of spin-ensembles by induced dephasing}
\par\end{center}{\large \par}

\section{Qubit dynamics within the rotating wave approximation}

Let us consider the Hamiltonian of Eq. (1) of the main article within
the rotating wave approximation, i.e. \begin{align}
H_{RW}=\omega_{S}S^{z}+\omega_{I}\sum_{k}I_{k}^{z}+\frac{J}{2}\sum_{k}\left[S^{+}I_{k}^{-}+S^{-}I_{k}^{+}\right].\label{hrt}\end{align}
 The total Hamiltonian has a $2\times2$ block-diagonal structure,
each block having a definite value of $S^{z}+I^{z}$. Its time-evolution
operator has the form \begin{align}
U=\sum_{I=0}^{N/2}\sum_{M_{I}=-I}^{I}\mathrm{e}^{i\phi_{M_{I}}t}\left\lbrace \cos\Omega_{M_{I}}t\hat{\mathcal{I}}+i\sin\Omega_{M_{I}}t\left[\frac{\Delta}{\Omega_{M_{I}}}\hat{\sigma}^{z}+\frac{\tilde{J}_{M_{I}}}{\Omega_{M_{I}}}\hat{\sigma}^{x}\right]\right\rbrace \hat{\mathcal{P}}_{I},\label{trt}\end{align}
 where $M_{I}$ is the magnetic quantum number of $I$, $\widetilde{J}_{M_{I}}=J\sqrt{(I-M_{I})(I+M_{I}+1)}$,
$\Delta=(\omega_{S}-\omega_{I})/2$ and $\Omega_{M_{I}}=\sqrt{\tilde{J}_{M_{I}}^{2}+\Delta^{2}}.$
$\hat{\sigma}$'s are the Pauli operators in the basis of the $2\times2$
block diagonals of $H_{RW}$. The projection operator $\hat{\mathcal{P}}_{I}$
corresponds to each bath-spin sector. The phase $\phi_{M_{I}}$ corresponding
to each block does not contribute to the dynamics, if the initial
state is diagonal in the $S^{z}+I^{z}$ basis.

The initial density matrix in the basis of $|\frac{1}{2};M_{I}\rangle$,
$|-\frac{1}{2},M_{I}+1\rangle$ is given by \begin{align}
\rho_{M_{I}}=\epsilon_{I}^{\frac{N}{2}+M_{I}}(1-\epsilon_{I}(0))^{\frac{N}{2}-(M_{I}+1)}\left[\begin{array}{cc}
\epsilon_{S}(0)(1-\epsilon_{I}(0)) & 0\\
0 & \epsilon_{I}(0)(1-\epsilon_{S}(0))\end{array}\right],\label{rim}\end{align}
 and the diagonal element evolutions are\begin{align}
\rho_{M_{I}}^{\left\{ \begin{array}{c}
ee\\
gg\end{array}\right\} }(t)=\left[1-\frac{\tilde{J}_{M_{I}}^{2}}{\Omega_{M_{I}}^{2}}\sin^{2}\Omega_{M_{I}}t\right]\rho_{M_{I}}^{\left\{ \begin{array}{c}
ee\\
gg\end{array}\right\} }+\frac{\tilde{J}_{M_{I}}^{2}}{\Omega_{M_{I}}^{2}}\sin^{2}\Omega_{M_{I}}t\rho_{M_{I}}^{\left\{ \begin{array}{c}
gg\\
ee\end{array}\right\} }.\label{etrt}\end{align}
 Complete exchange of polarization between $I$ to $S$, i.e. $\epsilon_{S}(\tau)=\epsilon_{I}(0)$,
is only possible at times $\tau$. This coherent exchange is destroyed
in presence of off-resonant fields ($\Delta\ne0$). In general, the
coherent exchange is controlled by the transfer coefficients \begin{align}
P_{+-}(n)=\frac{n^{2}J^{2}}{n^{2}J^{2}+(\omega_{S}-\omega_{I})^{2}},~~~1\le n=\sqrt{(I-M_{I})(I+M_{I}+1)}\le\sqrt{N},\end{align}
 and the spread of the incommensurate frequencies $\Omega_{M_{I}}$.

\subsection{Counter-rotating terms}

When the interaction Hamiltonian has counter-rotating ($CR$) terms
only, i.e., \begin{align}
H_{CR}=\omega_{S}S^{z}+\omega_{I}\sum_{k}I_{k}^{z}+\frac{J}{2}\sum_{k}\left[S^{+}I_{k}^{+}+S^{-}I_{k}^{-}\right],\label{hcr}\end{align}
 then $S^{z}+I^{z}$ is no longer a conserved quantity. Yet, the Hamiltonian
can be expressed in a block-diagonal form in the basis of $|\frac{1}{2};M_{I}\rangle$,
$|-\frac{1}{2},M_{I}-1\rangle$ . Similarly to the above analysis,
the time-evolution of the total system exchange energy between $S$
and $I$ determined by the factor\begin{align}
P_{++}(n)=\frac{n^{2}J^{2}}{n^{2}J^{2}+(\omega_{S}+\omega_{I})^{2}}.\end{align}

\section{Repeated measurements}

A measurement on the system which erases its off-diagonal elements
while keeping populations unchanged is a nearly ideal quantum-non-demolition
(QND) measurement. This measurement projects the system's state onto
its energy eigenbasis. In the present case this is also equivalent
to projecting also the $I$-spins on their (total) $I_{z}$ basis.
For the $RW$ and $CR$ interaction dynamics, the projective measurement
can be performed individually in each subspace of $S^{z}+I^{z}$.
By rewriting the initial density matrix in Eq. (\ref{rim}) as $\rho_{M_{I}}=\mathrm{Tr}[\rho_{M_{I}}]\left[\begin{array}{cc}
x_{M_{I}} & 0\\
0 & 1-x_{M_{I}}\end{array}\right]$ and accounting that off-diagonal elements are erased by each projective
measurement, if $n$ of them are performed at time-intervals $\tau$,
then \begin{align}
x_{M_{I}}(t=n\tau)=f_{1M_{I}}^{n}(\tau)x_{M_{I}}+f_{2M_{I}}(\tau)\sum_{m=0}^{n}f_{1M_{I}}^{m}(\tau),\end{align}
 where $f_{1}(t)=1-\frac{2\tilde{J}_{M_{I}}^{2}}{\Omega_{M_{I}}^{2}}\sin^{2}\Omega_{M_{I}}t$
and $f_{2}(t)=\frac{\tilde{J}_{M_{I}}^{2}}{\Omega_{M_{I}}^{2}}\sin^{2}\Omega_{M_{I}}t.$
In the limit of $n\rightarrow\infty$, $x_{M_{I}}(t)=\frac{f_{2M_{I}}(\tau)}{1-f_{1M_{I}}(\tau)}=\frac{1}{2}$
and the density matrix $\rho_{M_{I}}^{qe}=\rho_{M_{I}}(t\rightarrow\infty)=\mathrm{Tr}[\rho_{M_{I}}]\left[\begin{array}{cc}
\frac{1}{2} & 0\\
0 & \frac{1}{2}\end{array}\right].$ Hence the total density matrix (qubit + bath) commutes with the Hamiltonian
$H_{RW}$. A similar equilibrium state can be found for the $H_{CR}$.

\section{Steady-state values of the qubit populations}

The time-dependent population of the $S$ spin can be found by tracing
over the bath degrees of freedom obtaining\begin{equation}
\epsilon_{S}(t)=\epsilon_{S}(0)+\frac{\epsilon_{I}(0)-\epsilon_{S}(0)}{1-\epsilon_{I}(0)}F(t),\end{equation}
 where $F(t)=\sum_{I,M_{I}}W_{M_{I}}^{I}\sin^{2}\Omega_{M_{I}}t$
and $0\le F(t)\le1.$The weight function $W_{M_{I}}^{I}$ is given
by \begin{equation}
W_{M_{I}}^{I}=\lambda_{I}\epsilon_{I}(0)^{\frac{N}{2}+M_{I}}(1-\epsilon_{I}(0)_{I}^{\frac{N}{2}-M_{I}})J_{M_{I}}^{2}/\Omega_{M_{I}}^{2},~~0\le W_{M_{I}}^{I}\le1,\label{epst}\end{equation}
 where $\lambda_{I}=\frac{N!}{\left(N/2+I\right)!\left(N/2-I\right)!}\frac{2I+1}{N/2+I+1}$.
By performing repeated measurements as in the above section, the final
$S$-spin polarization within $RW$ approximation is given by \begin{align}
\epsilon_{S}(t=n\tau)=\epsilon_{S}(0)+\frac{\epsilon_{I}(0)-\epsilon_{S}(0)}{1-\epsilon_{S}(0)}F_{n}(\tau),\end{align}
 where \begin{align}
F_{n}(\tau) & =\frac{1}{2(\epsilon_{I}(0)-\epsilon_{S}(0))}\sum_{I,M_{I}}W_{M_{I}}^{I}\big[\left\lbrace \epsilon_{I}(0)(1-\epsilon_{s}(0))+\epsilon_{S}(0)(1-\epsilon_{I}(0))\right\rbrace (1-f(\tau))\sum_{m=0}^{n-1}f^{m}(\tau)\nonumber \\
 & -2\epsilon_{S}(0)(1-\epsilon_{I}(0))(1-f^{n}(\tau))\big].\end{align}
 In the above equation $f(\tau)=\cos2\Omega_{M_{I}}\tau$. After few
measurements $f^{n}(\tau)$ becomes negligibly small and $\sum_{m=0}^{n-1}f^{m}(\tau)\sim\frac{1}{1-f(\tau)}$.
Then, a steady state value is attained for any bath size $N$, given
by \begin{equation}
\epsilon_{S}^{qe}=\epsilon_{S}(0)+\frac{\epsilon_{I}(0)-\epsilon_{S}(0)}{2(1-\epsilon_{I}(0))}\left[1-\sum_{I}\lambda_{I}\epsilon_{I}(0)^{\frac{N}{2}+I}(1-\epsilon_{I}(0))^{\frac{N}{2}-I}\right].\label{eqrt}\end{equation}
 The equilibrium value for, $N=3$, can be evaluated from the above
equation, giving \begin{align}
\epsilon_{S}^{qe}=\epsilon_{S}(0)+\frac{1}{2}\left\lbrace (\epsilon_{I}(0)-\epsilon_{S}(0))[1+\epsilon_{I}(0)(1-\epsilon_{I}(0))]\right\rbrace .\end{align}
 Since $0\le\epsilon_{S}(0),\epsilon_{I}(0)\le1$, there will always
be a gain in polarization $\epsilon_{S}^{qe}>\epsilon_{S}(0)$, if
$\epsilon_{I}(0)>\epsilon_{S}(0)$. If the interaction Hamiltonian
has counter-rotating terms only, the projective equilibrium value
is given by \begin{align}
[\epsilon_{S}^{qe}]_{CR}=\epsilon_{S}(0)+\frac{1}{2}\left\lbrace (1-\epsilon_{I}(0)-\epsilon_{S}(0))[1+\epsilon_{I}(0)(1-\epsilon_{I}(0))]\right\rbrace .\end{align}
 Thus the counter-rotating terms take $\epsilon_{S}(0)$ close to
$1-\epsilon_{I}(0)$. \\
 \\
 \textbf{\large Large $N$ limit}{\large{} }\\
{\large{} }\\
{\large{} For large $N$ one can attain the maximum achievable
polarization transfer from the bath spins to the qubit (using $RW$
terms only). This saturation value, $\epsilon_{S}^{qe}\simeq\epsilon_{S}(0)+\frac{\epsilon_{I}(0)-\epsilon_{S}(0)}{2(1-\epsilon_{I}(0))},$
is reached for any value of the magnetic fields, $\omega_{S}$ and
$\omega_{I}$.}%
\begin{figure}[tbh]
\begin{centering}
\includegraphics[width=3.7313in,height=3.7885in]{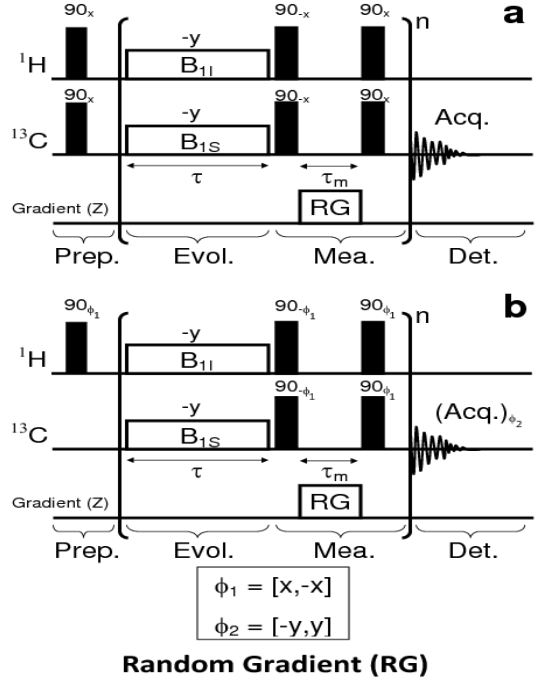} 
\par\end{centering}

\caption{\textbf{NMR pulse sequence.} (a) Both kinds of spins are excited for
the initial condition (Fig. 1, main text). (b) Based on the phase
cycling of pulses and the acquisition process, only the $I$ spins
are effectively excited as initial conditions of this variant (Fig.
2, main text). The time length of the $90{{}^{o}}$ pulses were $8\operatorname{\mu s}$
for the $I$ spins ($^{1}$H) and $15\operatorname{\mu s}$ for $S$
($^{13}$C). We used for the random intensities of the gradient values
in the range $\left\vert \Delta B_{z}\right\vert \leq30\operatorname{G}\operatorname{cm}$,
where the length of the sample is approximately $1\operatorname{cm}$.
This leads to a variation of the precession frequency around the static
field of approximately $125\operatorname{kHz}$ for $^{1}$H and $32\operatorname{kHz}$
for $^{13}$C. The time duration for the gradients is $\tau_{\mathrm{m}}=100\operatorname{\mu s}$,
satisfying the condition $\Delta\omega_{h}\tau_{\mathrm{m}}=\gamma_{S}\Delta B_{z}h\tau_{\mathrm{m}}\gg2\pi$.}

\label{NMRseq} 
\end{figure}
{\large \par}

\section{Experimental description of the NMR simulator}

\subsection{Hamiltonian realization}

The experiment described in the main text involves one spin $S$ and
three bath spins $I$ in the presence of a static magnetic field $B_{0}$
in the $z$ direction. Considering that the differences of the Larmor
frequencies are far greater than the $J$-coupling between the spins,
the effective static Hamiltonian is given by \begin{equation}
H=\omega_{I0}I^{z}+\omega_{S0}S^{z}+JI^{z}S^{z},\label{NMRHamiltonian}\end{equation}
 where $I^{z}=\sum_{k}I_{k}^{z}$, the sum being over bath spins.
When radio-frequency (RF) magnetic fields $B_{1I}$ and $B_{1S}$
oscillating at frequencies $\omega_{I0}$ and $\omega_{S0}$ are applied,
the Hamiltonian in a doubly rotating frame, precessing with the respective
Larmor frequencies, becomes \cite{Ernst-1} \begin{equation}
H=\omega_{I}I^{y}+\omega_{S}S^{y}+JI^{z}S^{z},\end{equation}
 where $\omega_{I}=\gamma_{I}B_{1I}$ and $\omega_{S}=\gamma_{S}B_{1S}$.

We assume that the direction of the RF fields is $-y$ in this frame.
By making a change of variables for the spatial directions ($y\rightarrow z,\, z\rightarrow x$)
this Hamiltonian becomes equivalent to the one given in Eqs. (1) and
(2) of the main text, \begin{equation}
H=\omega_{I}I^{z}+\omega_{S}S^{z}+JI^{x}S^{x}.\label{eq:Hmain}\end{equation}
We would like to note that the presence of both $RW$ and $CR$ terms
in the above Hamiltonian is a direct consequence of the anisotropic
$I^{x}S^{x}$ coupling. The $\vec{I}\cdot\vec{S}$ or $S^{x}I^{x}+S^{y}I^{y}$
coupling, as found in many natural spin systems, would result only
in AZE cooling and quasi-steady state behavior discussed in the main
text. The QZE heating and the reheating effects would be absent, because
of the cancellation of the $CR$ terms.

\subsection{Pulse sequence with random gradients}

Figure \ref{NMRseq} is a schematic representation of the NMR pulse
sequence used for simulating the measurement-induced quantum dynamics
described in the main text. The preparation of the initial state,
shown in Fig. \ref{NMRseq}, consists of the following steps:

(i). $90{{}^{o}}$ RF pulses applied along the $x$ axis %
\footnote{The spatial directions are defined with respect to the doubly rotating
frame which precesses with the Larmor frequencies of the spins.%
}, exactly on-resonance with the respective Larmor frequencies of both
spin species. These pulses rotate the magnetizations from the initial
$z$ direction to the $-y$ direction. This is followed by the application
of on-resonance RF fields pointing to the $-y$ direction on both
kinds of spins, and the Hamiltonian (\ref{eq:Hmain}) is obtained
during time $\tau$.

(ii). In order to simulate the projective measurements, we rotate
the polarization back to the $z$ axis by $90{{}^{o}}$ pulses on
the $-x$ axis. At this point a static field gradient is applied during
time $\tau_{m}$. To simulate a projective measurement, the gradient
intensity is chosen randomly (see section \ref{sec:Gradient-Driven-Projective-Measurements}).

Thus the Hamiltonian generated by the field gradients in a slide of
the sample at a position $z$ on the axis parallel to the gradient
direction is given by $H_{M}^{z}(t)\propto\sum_{n}\left(\gamma_{S}\Delta B_{z}^{n}zS_{z}+\gamma_{I}\Delta B_{z}^{n}zI_{z}\right)\theta\left(t-n\tau\right)\theta\left(n\tau+\tau_{m}-t\right)$
, where $\Delta B_{z}^{n}$ is the gradient strength at the step $n$.
After each $\tau_{m}$ time, the ensemble average of the different
portion of the sample preserves the populations in the density matrix
but erases the quantum correlations between the $I$ and $S$ spins.
Additionally, the non-diagonal elements of the density matrix within
the blocks of constant total spin of $I$, the so-called zero quantum
coherences, are not erased, due to the full chemical equivalence of
the $I$ spins. 

(iii). Following this projection the polarizations are returned to
the $-y$ direction by $90{{}^{o}}$ pulses on the $x$ axis, and
the entire cycle is repeated as schematized in Fig. \ref{NMRseq}.
After $n$ cycles, we realize the acquisition of the $S_{y}$ magnetization
as if $n$ projective measurements were performed on $S$. As the
coherences between $I$ spins are not modified by the gradients, the
simulated measurements mimic measurements only on the $S$ spin.

(iv). Alternatively one can prepare an initial excitation on the $I$
spins only. For this purpose, during the preparation time, only the
$I$ spins are rotated to the $-y$ direction, keeping the $S$ magnetization
on the $z$ axis. By phase cycling we cancel out the initial contribution
of the $S$ spins (i.e., $\epsilon_{S}(0)=1/2$).

\subsection{Pulse sequence without random gradients}

In order to compare these evolutions with the free-evolution dynamics,
we repeated the experiment with the same pulse sequences but with
null gradients. This procedure gives the same time scales for the
relaxation and for other non-ideal features of the pulses, but does
not modify the free dynamics between measurements. Decoherence is
manifest by the reduction of the oscillation amplitude as time goes
on. However, the first oscillations are comparable for both procedures.
This provides the evidence that there is no polarization transfer
during the $\pi/2$ pulses. The polarization control by simulated
measurements discussed in the paper is only achieved when the random
gradients are introduced to dephase of the coherences (correlations)
between $S$ and $I$ spins (Eq. \ref{eq:gradientexp}).

\section{Gradient-Driven Projective Measurements in NMR\label{sec:Gradient-Driven-Projective-Measurements}}

The experimental simulations of repeated projective measurements were
achieved with \textit{random gradient fields}. A brief analysis of
this equivalence is given below.

We consider a $S$ spin in presence of a magnetic field on the $y$
axis whose Zeeman frequency is $\omega_{S}$. Its initial state is
$\rho_{S}=\exp(-\beta\omega_{0S}S^{z})/\mathcal{Z}\approx\left(\mathbf{1}-\beta\omega_{0S}S^{z}\right)/\mathcal{Z}$
in the high temperature limit, a typical condition in NMR experiments
\cite{Ernst-1}. Its evolution is given by\begin{equation}
\rho\left(\tau\right)=S^{z}\cos\left(\omega_{S}\tau\right)+S^{x}\sin\left(\omega_{S}\tau\right),\label{DM1spins}\end{equation}
 where we do not write the constant evolution of the identity operator
and the factor $-\beta\omega_{0S}/\mathcal{Z}$. By performing projective
measurements at time intervals $\tau$ on the energy eigenbasis of
$S^{z},$ $S^{x}\sin\left(\omega_{S}\tau\right)$ is erased. After
$n$ measurements $\rho_{n}\left(n\tau\right)=S^{z}\cos^{n}\left(\omega_{S}\tau\right).$
Applying a field gradient on the $z$ axis at time intervals $\tau$
during a time $\tau_{m}$ to simulate the measurements, a slice of
the sample at position $z$ is affected by the field $\Delta B_{z}z,$
where $\Delta B_{z}$ is the gradient value. The resulting evolution
of the slice\textasciiacute{}s density matrix (\ref{DM1spins}) is
$\rho_{z}\left(\tau\right)=S^{z}\cos\left(\omega_{S}\tau\right)+\left[S^{x}\cos\left[\Delta\omega_{z}\tau_{\mathrm{m}}\right]+S^{y}\sin\left(\Delta\omega_{z}\tau_{\mathrm{m}}\right)\right]\sin\left(\omega_{S}\tau\right),$
where $\Delta\omega_{z}=\gamma_{S}\Delta B_{z}z.$ Its ensemble average
within the sample length $h$ is then $\rho\left(\tau\right)=\int_{-h/2}^{h/2}\rho_{z}\left(\tau\right)\frac{\mathrm{d}z}{h}=S^{z}\cos\left(\omega_{S}\tau\right),$
assuming that $\Delta\omega_{h}\tau_{\mathrm{m}}=\gamma_{S}\Delta B_{z}h\tau_{\mathrm{m}}\gg2\pi$.

While this result represents a projective process for $n=1$, the
next step $n=2$ is no longer the outcome of two consecutive projections.
By calculating the density matrix of the portion $z$ of the sample
at this step and considering the ensemble average\begin{multline}
\rho\left(2\tau\right)=\int_{-h/2}^{h/2}\rho_{z}\left(2\tau\right)\frac{\mathrm{d}z}{h}=S^{z}\cos^{2}\left(\omega_{S}\tau\right)+S^{x}\sin\left(\omega_{S}\tau\right)\int_{-h/2}^{h/2}\left[\cos\left(\omega_{S}\tau\right)\cos^{2}\left(\Delta\omega_{z}\tau_{\mathrm{m}}\right)-\sin^{2}\left(\Delta\omega_{z}\tau_{\mathrm{m}}\right)\right]\frac{\mathrm{d}z}{h}.\label{eq:gradientexp}\end{multline}
 While the first term on the rhs is equivalent to the one obtained
by two projective measurements, but the second term is non-vanishing.
In general, correlations of multiple steps, of the form $\cos^{m}\left(\Delta\omega_{z}\tau_{\mathrm{m}}\right)\sin^{l}\left(\Delta\omega_{z}\tau_{\mathrm{m}}\right)$
with $l$ and $m$ even numbers, survive the ensemble average, leading
to non vanishing deviations from projective measurement. Therefore,
a field gradient by itself does not simulate repeated projective measurements.
By changing randomly the gradient intensities at consecutive steps,
we destroy correlations among different steps upon taking the ensemble
average. Thus only the term $S^{z}\cos^{n}\left(\omega_{S}\tau\right)$
survive the ensemble average, properly mimicking the projective measurements.

Similar results are obtained once considered the full $S$-$I$ system.
The Hamiltonian $H_{M}^{z}(t)$ described in the preceding section
dephases $S_{x,y}$ and $I_{x,y}$ terms in the density matrix when
the ensemble average on the position $z$ is taken. An ideal projective
measurement is performed when all spins along $z$ experiences a random
field. A single gradient gives a particular $z$-dependent phase to
spins along the sample. Only when this gradient is randomized further
many times such that the sum of the phases accumulated under each
gradient vanishes for each spin, one can realize an ideal projective
measurement as showed above. However, because the non-diagonal elements
of the density matrix within the blocks of constant total spin of
$I$ are not erased, due to the full chemical equivalence of the $I$
spins, the simulated measurements mimic measurements only on the $S$
spin.

\end{widetext}
\end{document}